\documentclass[aip,rsi,amsmath,amssymb,reprint]{revtex4-1}

\usepackage[T1]{fontenc}
\usepackage[utf8]{inputenc}
\usepackage[english]{babel}
\usepackage{times}
\usepackage[pdftex]{graphicx} 
\usepackage[usenames,dvipsnames,svgnames,table,rgb]{xcolor}
\usepackage{color}
\usepackage{setspace}
\usepackage{dcolumn}
\usepackage{longtable}
\usepackage{multirow}

\usepackage[normalem]{ulem} 

\addto\captionsenglish{}

\addto\captionsenglish{}

\usepackage{hyperref}
\hypersetup{%
  colorlinks=true,
  linkcolor=blue,
  urlcolor=blue,
  citecolor=blue,
  pdftitle={High energy-resolution x-ray spectroscopy at ultra-high dilution with spherically bent crystal analyzers of 0.5~m radius},%
  pdfauthor={M. Rovezzi et al.},%
  pdfsubject={},%
  pdfkeywords={crystal analyzer, x-ray spectrometer, x-ray absorption spectroscopy, x-ray emission spectroscopy, monochromator, wavelength-dispersive spectrometry}%
}

\begin{document}
\setlength{\parskip}{0pt} 
\preprint{AIP/123-QED}

\title{High energy-resolution x-ray spectroscopy at ultra-high dilution with spherically bent crystal analyzers of 0.5~m radius}
\author{Mauro~Rovezzi}
\email{mauro.rovezzi@esrf.eu}
\affiliation{European Synchrotron Radiation Facility, 71 avenue des Martyrs, CS 40220, 38043 Grenoble, France}
\author{Christophe~Lapras}
\affiliation{European Synchrotron Radiation Facility, 71 avenue des Martyrs, CS 40220, 38043 Grenoble, France}
\author{Alain~Manceau}
\affiliation{ISTerre, Universit{\'e} Grenoble Alpes, CNRS, CS 40700, 38058 Grenoble, France}
\author{Pieter~Glatzel}
\email{pieter.glatzel@esrf.eu}
\affiliation{European Synchrotron Radiation Facility, 71 avenue des Martyrs, CS 40220, 38043 Grenoble, France}
\author{Roberto~Verbeni}
\email{roberto.verbeni@esrf.eu}
\affiliation{European Synchrotron Radiation Facility, 71 avenue des Martyrs, CS 40220, 38043 Grenoble, France}
\date{\today}

\begin{abstract}
  We present the development, manufacturing and performance of spherically bent crystal analyzers (SBCAs) of 100~mm diameter and 0.5~m bending radius. The elastic strain in the crystal wafer is partially released by a ``strip-bent'' method where the crystal wafer is cut in strips prior to the bending and the anodic bonding process. Compared to standard 1~m SBCAs, a gain in intensity is obtained without loss of energy resolution. The gain ranges between 2.5 and 4.5, depending on the experimental conditions and the width of the emission line measured. This reduces the acquisition times required to perform high energy-resolution x-ray absorption and emission spectroscopy on ultra-dilute species, accessing concentrations of the element of interest down to, or below, the ppm (ng/mg) level.\\
\end{abstract}


\maketitle

\section{Introduction}
\label{sec:introduction}

Fluorescence detected x-ray absorption spectroscopy (XAS) on samples with low absorber concentration is usual\-ly measured with multi- or single-element solid state detectors (SSDs), which have an intrinsic energy resolution $\Delta$E/E~$>$1$\times$10$^{-2}$ limited by statistical fluctuations in the number of electron-hole pairs produced by the radiation~\cite{Fano:1947_PR} and the electronic noise. In addition, SSDs have a counting rate per detector element typically lower than 1~MHz if the photons are uniformly distributed in time, which may reduce significantly when the x-rays arrive as short and intense pulses (\emph{e.g.} timing modes at synchrotron radiation facilities). For dilute samples, poor energy resolution is a limitation because the weak emission signal from the element of interest is not fully resolved from the stronger elastic and inelastic scattering signal from the matrix. Elements other than the desired absorber contained in the matrix may give rise to additional fluorescence lines and thus effectively reduce the available maximum count rate for the fluorescence line of the element of interest. These two limitations can be overcome by using a spectrometer. An optical element is inserted between the sample and the detector, which acts as collector and mono-/poly-chromator. For the hard x-ray energy range, a crystal analyzer is generally used as optical element. It can be used either as Bragg (reflection) or Laue (transmission) optics. It provides an energy resolution of typically $\Delta$E/E~=~1--50$\times$10$^{-5}$ (0.1--5~eV bandwidth at 10~keV) and thus selects the fluorescence line of interest that is registered by the detector.

An instrumental energy bandwidth on the order or below the natural linewidth of the fluorescence line measured (0.5--5~eV) allows for a sharpening effect of the spectral features.~\cite{Hamalainen:1991_PRL,DeGroot:2002_PRB,Glatzel:2012_JESRP} This type of measurement is called high energy-resolution fluorescence-detected (HERFD), partial fluorescence yield (PFY) or high resolution (HR) XAS. A crystal analyzer allows also performing x-ray emission spectroscopy (XES) and resonant inelastic x-ray scattering (RIXS), which have received considerable attention because of a better theoretical understanding and the advent of x-ray free electron lasers. For a recent topical review, see Refs.~\onlinecite{Rovezzi:2014_SST,Glatzel:2016_book} plus references therein. Measurements with crystal analyzers are increasingly embraced by researchers in the applied science who seek to study dilute and/or radiation sensitive samples. To fulfill this goal, it is necessary to increase also the lu\-mi\-no\-si\-ty of the spectrometers, ideally without degrading the energy resolution.

Many of the spectrometers dedicated to HERFD-XAS, XES and RIXS are based on curved analyzers arranged in a Rowland-circle geometry,~\cite{Rowland:1882_PhilMag} a Bragg reflection geometry which allows simultaneous focusing and energy analysis of a fluorescence source (4$\pi$ divergence). In such geometry, the sample, crystal analyzer and detector are typically positioned on a circle, the Rowland circle. The crystal analyzer is usually designed either to work in a point-to-point scanning configuration or in a dispersive one. In the first arrangement, the crystal analyzer acts as a monochromator. A small energy bandwidth ($<$5~eV) satisfies Bragg's diffraction condition over the whole crystal area and is focused on the detector which acts as a photon counter (\emph{e.g.} avalanche photodiode, proportional counter). The energy is selected by moving the spectrometer components on the Rowland circle, according to the mechanical design of the spectrometer. In the second arrangement, the crystal analyzer acts as a polychromator, either by design (\emph{e.g.} von H\'amos) or by moving it out of the Rowland circle. A relatively large bandwidth ($<$100~eV) is dispersed over an area on a position-sensitive detector (\emph{e.g.} charge-coupled device, pixel detector). Each point on the detector along the energy dispersive direction corresponds to a different energy. Thus, only a small fraction of the crystal surface contributes to a given ener\-gy.

Point-to-point scanning spectrometers are usually based on spherically bent crystal analyzers (SBCAs). Such instruments are built not only at synchrotron radiation facilities~\cite{Shvydko:2004_book_ch6,Welter:2005_JSR,Fister:2006_RSI,Hazemann:2009_JSR,Verbeni:2009_JSR,Kleymenov:2011_RSI,Llorens:2012_RSI,Shvydko:2012_JESRP,Sokaras:2013_RSI,Kvashnina:2016_JSR}, but also at plasma physics facilities~\cite{Bitter:2003_RSI,Kugland:2011_JINST} and in the laboratory.~\cite{Seidler:2014_RSI,Mortensen:2016_conf,Seidler:2016_conf} SBCAs are employed near back-scattering conditions at a Bragg reflection angle $\ge$65$^\circ$. This allows minimizing the Johann error~\cite{Johann:1931_ZP} caused by the deviation of the crystal surface out of the Rowland circle~\cite{Bergmann:1998_conf}, because the bending radius of the crystal wafer is twice the radius of the Rowland circle. When used in back-scattering configuration, SBCAs give the best compromise in terms of energy-resolution, focal size, collection efficiency and manufacturing.~\cite{Wittry:1992_JAP}

State-of-the-art technology for the production of SBCAs is based on elastically bending high quality thin crystal wafers, such as silicon, germanium, or quartz, on spherical substrates. The crystal is forced on a precisely machined concave glass or metallic substrate by applying a uniform pressure via a convex die (anvil). Then the bonding between the crystal and the substrate is performed either by epoxy gluing or via anodic bonding.~\cite{Wallis:1969_JAP,Verbeni:2005_JPCS,Collart:2005_JSR} Anodic bonding requires the control of additional parameters including temperature and voltage. It also requires the use of borosilicate glass as substrate and is presently limited to silicon (Si). The most used crystal thickness is between 300~$\mu$m and 500~$\mu$m, which gives the best results for bending radii $\ge$1~m for a diameter of 100~mm.

Standard circular SBCAs of 1~m bending radius and 100~mm diameter, have an energy bandwidth below the natural x-ray emission line width but the solid angle collected is as low as $\approx$0.008~sr, less than 0.1\% of the whole emitted fluorescence signal. In order to increase the detection efficiency, the collected solid angle must be increased. The first option is to increase the number of crystal analyzers. For a scanning spectrometer this is extremely difficult from a mechanical design point of view, because each crystal analyzer defines its own Rowland circle and must be moved accordingly. Current instruments have usually one row with five to seven SBCAs. The second option consists in reducing the distance to the sample, \emph{i.e.} reducing the bending radius. A factor four in the collected solid angle is gained with SBCAs when moving from 1~m to 0.5~m.

SBCAs at 0.5~m bending radius are challenging to manufacture. Even with thin wafers of 150 $\mu$m, the crystal has to accommodate a high strain when forced on the sphe\-ri\-cal substrate. One solution is to reduce the diameter of the crystal analyzer, but this decreases the collected solid angle. Alternatively, a plastic deformation of the crystal may be performed, as for example in the hot plastic deformation method.~\cite{Okuda:2008_JAC} However, this may introduce high mosaicity and degrade con\-si\-de\-ra\-bly the energy resolution.

We present here a simple and effective approach, which consists in cutting a thin crystal wafer in strips prior to the bending and the anodic bonding process. The method is described in \S\ref{sec:exp}. Optical performances and an example of application are presented in \S\ref{sec:res}.

\section{Methods}
\label{sec:exp}
%
\subsection{Analyzers production}
\label{sec:exp_prod}

The analyzers are routinely produced at the ESRF Cry\-stal Analyzer Laboratory (CAL). The method described here is optimized for the anodic bonding of Si. The method may be equally applicable for any material if the crystal is fixed to the substrate by epoxy gluing instead of anodic bonding. The standard approach employed for 1~m SBCAs, which consists of elastically bending a uniform wafer, is not effective for 0.5~m SBCAs. Most of the time, the thin 100~mm diameter Si wafer cracks when applying the pressure during the temperature ramp. In most cases, the wafer or the underlying glass substrate simply explode under the internal strain.

\begin{figure}[!htb]
  \centering
  \includegraphics[width=0.49\textwidth]{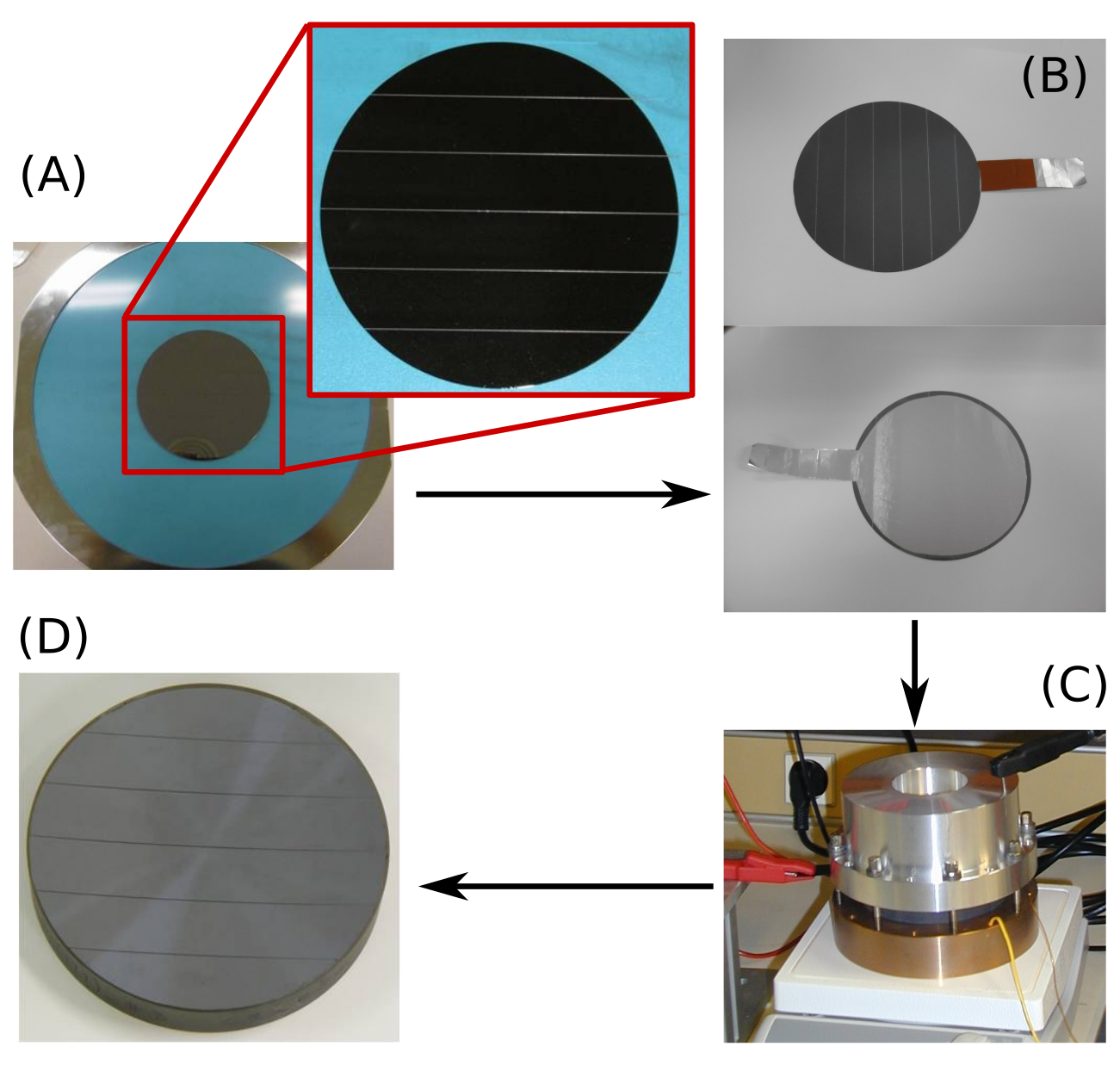}
  \caption{(Color online) ``Strip-bent'' procedure for the fabrication of 0.5~m SBCAs (\emph{cf.} main text).}
  \label{fig:procedure}
\end{figure}

\begin{figure}[!htb]
  \centering
  \includegraphics[width=0.49\textwidth]{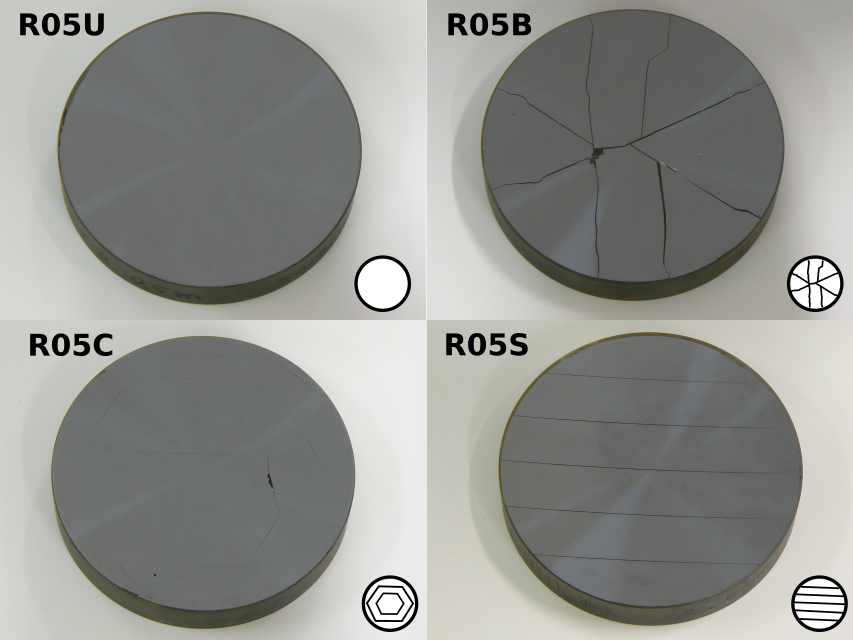}
  \caption{(Color online) ESRF 0.5~m Si(111) SBCAs employed for the x-ray characterization: uniform (R05U), badly cracked (R05B), cracked along cleavage planes (R05C) and strip-bent (R05S).}
  \label{fig:tested}
\end{figure}

The production yield is increased by employing the ``strip-bent'' procedure showed in Fig.~\ref{fig:procedure}: (A) a thin crystal wafer (150 $\mu$m thick) is cut in strips of 15~mm width; (B) a thin aluminum tape is then used to keep the strips together and act as electrode in the anodic bonding press; (C) the anodic bonding process~\cite{Verbeni:2005_JPCS} is performed on the polished side without tape by bending the crystal at $\approx$250~$^\circ$C and applying a voltage of 1500~V; (D) the tape is removed and the surface cleaned. The quality of the crystal surface is not affected by the use of the conductive tape. Four representative 0.5~m SBCAs are presented here, as shown in Fig.~\ref{fig:tested}: uniform (R05U), badly cracked (R05B), cracked along cleavage planes (R05C) and strip-bent (R05S). Crystals R05U, R05B and R05C were obtained in the early stage of the process and each has a specific type of me\-cha\-ni\-cal defect which cannot be reproduced exactly. In contrast, R05S can be reproduced in series. The quality of the ana\-ly\-zers is routinely tested with x-rays. Commercial SBCAs provided by companies including XRS TECH LLC \cite{_fn:XRSTECH} in the USA (XRS) and Saint-Gobain Crystals \cite{_fn:SG} in France were also tested in order to evaluate the quality of the process. Results obtained on strip-bent analyzers produced by XRS via anodic bonding \cite{xrstech:2016_patent} are equivalent to ours (not shown).

\subsection{X-ray characterization}
\label{sec:exp_tests}

\begin{figure}[!htb]
  \centering
  \includegraphics[width=0.49\textwidth]{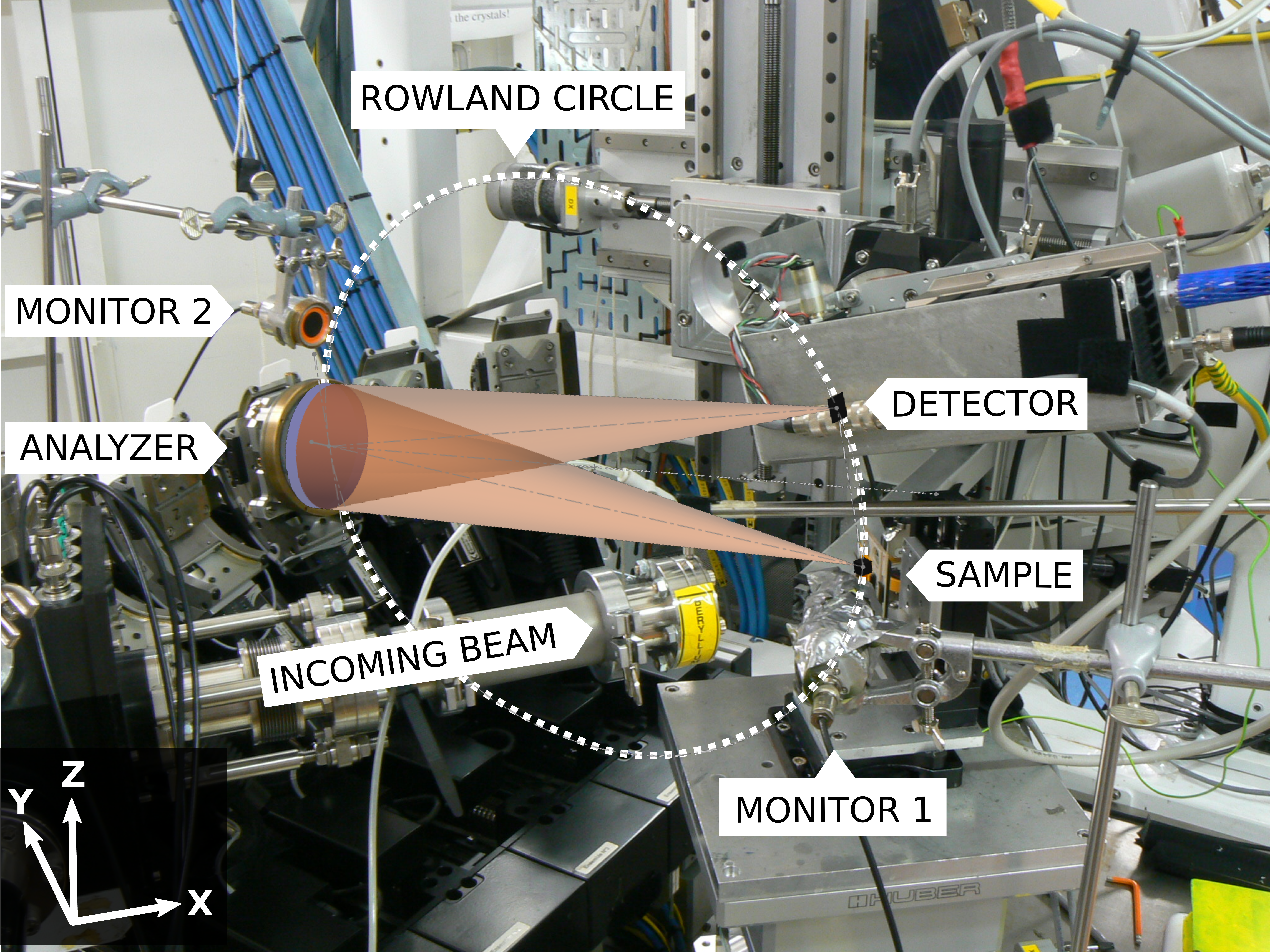}
  \caption{(Color online) Geometry and layout of the experiment: sample, crystal analyzer and detector are positioned on a vertical Rowland circle (dashed withe circle). The spectrometer was rotated in the horizontal plane toward back-scattering with an angle of 55$^\circ$ between the incoming beam and the center of the analyzer. The Helium bag has been taken out for clarity.}
  \label{fig:expsetup}
\end{figure}

The characterization with x-rays was conducted at the beamline ID26 of the European Synchrotron Radiation Facility (ESRF) with a point-to-point scanning spectrometer in a vertical Rowland circle geometry.~\cite{Glatzel:2009_CT,Kleymenov:2011_RSI} The experimental setup consists of an incoming beam monochromator equipped with a pair of cryogenically cooled flat crystals arranged in double crystal parallel configuration. The characterization of the crystal analyzers was performed with the Si(311) monochromator, giving an energy resolution of $\approx$4$\times$10$^{-5}$, corresponding to an incoming beam energy bandwidth of $\approx$0.28~eV at 7~keV. For ultra-high dilution applications the Si(111) monochromator was used, which had an energy resolution of $\approx$1.43$\times$10$^{-4}$, corresponding to an incoming beam energy bandwidth of $\approx$1~eV at 7~keV. The beam was focused on the sample via two Pd-coated Si mirrors in Kirkpatrick-Baez configuration, giving a beam size of $\approx$50$\times$500~$\mu$m$^2$ (vertical $\times$ horizontal). The incoming beam slits at $\approx$1~m before the sample were open to 100$\times$800~$\mu$m$^2$. To reduce air absorption in the beam path, a plastic bag with 25~$\mu$m thick Kapton\textsuperscript{\textregistered} (polyimide) windows was filled with He and kept under a flow of 4 l/h. To increase the elastic scattering intensity, the spectrometer was rotated in the horizontal plane toward back-scattering with an angle of 55$^\circ$ between the incoming beam and the center of the analyzer. The sample was rotated at 75$^\circ$ incidence angle. The tested SBCAs were always mounted on the same stage of the spectrometer with the center of the crystal aligned at the same height as the beam. To precisely compare the scattered intensities two Si photo-diodes were used as beam monitors. A sketch of the geometry and the layout (without the He bag) of the experiment is shown in Fig.~\ref{fig:expsetup}.

Two detectors were employed for these tests: 1) a 120 $\mu$m thick avalanche Si photo-diode of 10$\times$10~mm$^2$ active area. This is a single element photon counting detector and does not have spatial resolution. The focal scans with this detector were performed using vertical slits in front of the detector closed to 100~$\mu$m. 2) a MAXIPIX~\cite{Ponchut:2011_conf} 1$\times$1 with remote head of 14$\times$14~mm$^2$ active area with 256$\times$256 square pixels of 55~$\mu$m size. The second detector allows imaging the focus upon changing the energy of the incoming beam. In addition, by moving the detector outside the Rowland circle, it is possible to collect an image of the scattered intensity over the analyzer surface. The focal properties of the tested SBCAs were investigated by combining these two detectors. The measured focal size did not differ significantly from what is expected from ray tracing simulations. Imaging applications or the in-focus dispersive correction~\cite{Huotari:2005_JSR} for improving the ener\-gy resolution of the crystal analyzer are beyond the scope of this manuscript.

The crystal analyzers were characterized by the full width at half maximum (FWHM) and intensity of the reflectivity curve, also called re\-so\-lu\-tion function or elastic peak. It was measured by scanning the incident photon energy across the range of energies corresponding to the crystal analyzer reflection, given by the Bragg angle $\theta_{\rm B}$, the $d$-spacing and the order of reflection. The analyzer was kept fixed and a SIGRADUR\textsuperscript{\textregistered} (glassy carbon, $\approx$1~mm thick) sample was used as scatterer. For each SBCA, a given orientation, reflection and $\theta_{\rm B}$, the testing procedure consisted of: 1) geometrically aligning the sample, the center of crystal surface and the detector slit on the Rowland circle with a precision of 100~$\mu$m. 2) adjusting the alignment of the analyzer on the Rowland circle with nominal radius 0.5~m or 1~m, using a fluorescence line and by rocking scans in the two orthogonal directions: meridional ($\theta_{\rm B}$ scan) and sagittal ($\chi$ scan). 3) optimizing the bending radius parameter, R, by finding the minimum of a second order polynomial function, $f({\rm R}) = {\rm MAX} * {\rm SUM} / {\rm FWHM}$, where MAX, SUM and FWHM are, respectively, the maximum, sum of counts and full width at half maximum parameters obtained by the fit of the elastic peak with an asymmetric Pseudo-Voigt profile (a linear combination of a Gaussian and a Lorentzian function) \cite{_fn:sloth}. The bending radius optimization was done in a range of R~$\pm$~20~mm. 4) extracting the energy bandwidth (FWHM) from the fit of the elastic peak at the best radius. 4$^\prime$) optional, repeating an elastic peak scan and extracting the energy bandwidth at the nominal radius. 5) extracting the vertical FWHM of the focal point on the Rowland circle via a scan with a 100~$\mu$m slit in front of the detector and using an energy broad source, the fluorescence line used for the initial alignment. All data were divided point by point by a monitor signal collected $\approx$50~mm from the sample at $\approx$55$^\circ$ scattering with a photodiode (``monitor 1'' shown in Fig.~\ref{fig:expsetup}). This monitor was used to normalize the signal of the crystal analyzer to the incoming beam flux fluctuations and the scattered or fluorescence signal from the sample. This normalization allows for quantification of the relative scattered intensities from the SBCAs.

The Si(111) SBCAs presented here were tested in the energy range 6--11 keV, using the Si(333), Si(444) and Si(555) reflections. No major differences in terms of energy resolution and focus were found among these three orders of reflection. For this reason, only the results on the Si(555) reflection are reported. The alignment at this reflection was done using the Hg L$_3$-M$_5$ (L$\alpha_1$) line of $\alpha$-HgS (red cinnabar) powder diluted to 10\% in cellulose (energy~9989 eV, $\theta_{\rm B}$~81.73$^\circ$). The elastic peaks were collected at $\theta_{\rm B}$ of 82$^\circ$ and 75$^\circ$. These two angles were chosen because the use of two angles allows one to separate the intrinsic from the geometrical contributions in the measured energy resolution, as explained in the following. The influence from intrinsic and all geometrical contributions varies between 82$^\circ$ (mainly intrinsic) and 75$^\circ$ (mainly geometrical). In addition, most of the point-to-point scanning spectrometers work between 89$^\circ$ and 65$^\circ$. It is useful to show how the performance of the analyzers degrades going off back scattering.

The azimuthal orientation of the crystal was also tested to ensure that two reflections were not diffracting at the same time in any measurement. For example, this unfortunate situation may occur with the Si(555) and Si(751) reflections which have identical $d$-spacing and an angle of $\approx$29.9$^\circ$ between the two planes. For a Bragg angle of 75$^\circ$, the [555] and [751] normal axes make an angle of $\approx$15$^\circ$ with the incoming beam, therefore there is one azimuthal orientation for which the two planes diffract simultaneously. The test consisted in verifying that neither the energy bandwidth nor the reflected intensity changed when the azimuthal angle was varied.

The intrinsic broadening depends on the Darwin width,~\cite{Batterman:1964_RMP} the extinction length and the elastic strain caused by the bending.~\cite{Honkanen:2013_JSR,SanchezDelRio:2015_JAC} This broadening is the same for each point on the crystal surface. In contrast, the geometric contributions depend on the size and shape of the reflecting surface. Any angular divergence, $\Delta\theta$, contributes to the energy resolution via the differential form of Bragg's law: $\Delta$E/E~=~$\Delta\theta$/$\tan(\theta)$. For a perfect sphere, among all geometric contributions,~\cite{Bergmann:1998_conf,Suortti:1999_JSR} the most important are the Johann error and the fact that the crystal extends out of the Rowland plane in the sagittal direction. In this respect, the source volume plays an important role, especially in the dispersive direction. In our experimental conditions the beam size was sufficiently small such that the source volume did not influence the final measured energy resolution. A geometric contribution is also given by the fi\-gu\-re error, the fact that the crystal wafer may not form a perfect sphere when fabricated. This contribution is expected to be negligible, because the anodic bonding is a very strong bonding and, if the crystal surface and substrate are clean from impurities, the final shape of the analyzer is dictated by the quality of the underlying substrate, whose shape is machined for optical quality. In contrast, the flatness of the crystal wafer face which is bonded to the substrate is an additional source of geometrical error. Our optical measurements found 1--2~$\mu$m peak-to-valley over a diameter of 90~mm.

\subsection{Dynamical diffraction and ray tracing simulations}
\label{sec:exp_sims}

In order to fully understand the experimental results, a virtual experiment was simulated. For each of the two bending radii, 1~m and 0.5~m, the simulations were performed at two angles of the Si(555) reflection, 82$^\circ$ and 75$^\circ$. The study consisted in: 1) calculating the diffraction profile of the spherically bent crystal; 2) performing a ray tracing simulation.

The multi-lamellar (ML) method as implemented in the {\sc crystal} module~\cite{SanchezDelRio:2015_JAC}, part of {\sc xop}~\cite{SanchezdelRio:2011_SPIE}, was employed for calculating the reflectivity curve of the bent crystal and was used as input for the ray tracing simulations. This method is robust and well tested, but calculates only the perpendicular contribution to the strain field caused by the crystal bending. Recently, it was shown~\cite{Honkanen:2013_JSR,Honkanen:2014_JSR,Honkanen:2016_JAC} that a more realistic diffraction profile of a spherically bent crystal is obtained including also the angular contribution to the strain field which arises from the compression of the crystal on the substrate. This contribution broadens considerably the reflectivity of the crystal. The new approach calculates the perpendicular strain field integrating the one-dimensional Takagi-Taupin (TT) equations and the angular strain field via an analytical model. Both results are reported for comparison and referred to as vertical and angular for simplicity. The calculations of the diffraction profiles were performed for a crystal of 150~$\mu$m thickness.
 
The ray tracing (RT) part was performed using the {\sc shadow3}~\cite{SanchezdelRio:2011_JSR} code. The experimental source was modeled with a geometric source of elliptical shape of 100$\times$800~$\mu$m$^2$ and conical divergence. The divergence of the source was adjusted for each configuration (\emph{i.e.}, 82$^\circ$ and 75$^\circ$) in such a way to illuminate the whole crystal surface with 10$^5$ rays. The virtual experiment was performed by scanning the energy of the monochromatic source in steps of 0.05~eV and integrating the intensity of all the rays reaching the detector at each step.

\section{Results and discussion}
\label{sec:res}

\subsection{Optical performances}
\label{sec:performances}

\begin{figure}[!htb]
  \centering
  \includegraphics[width=0.49\textwidth]{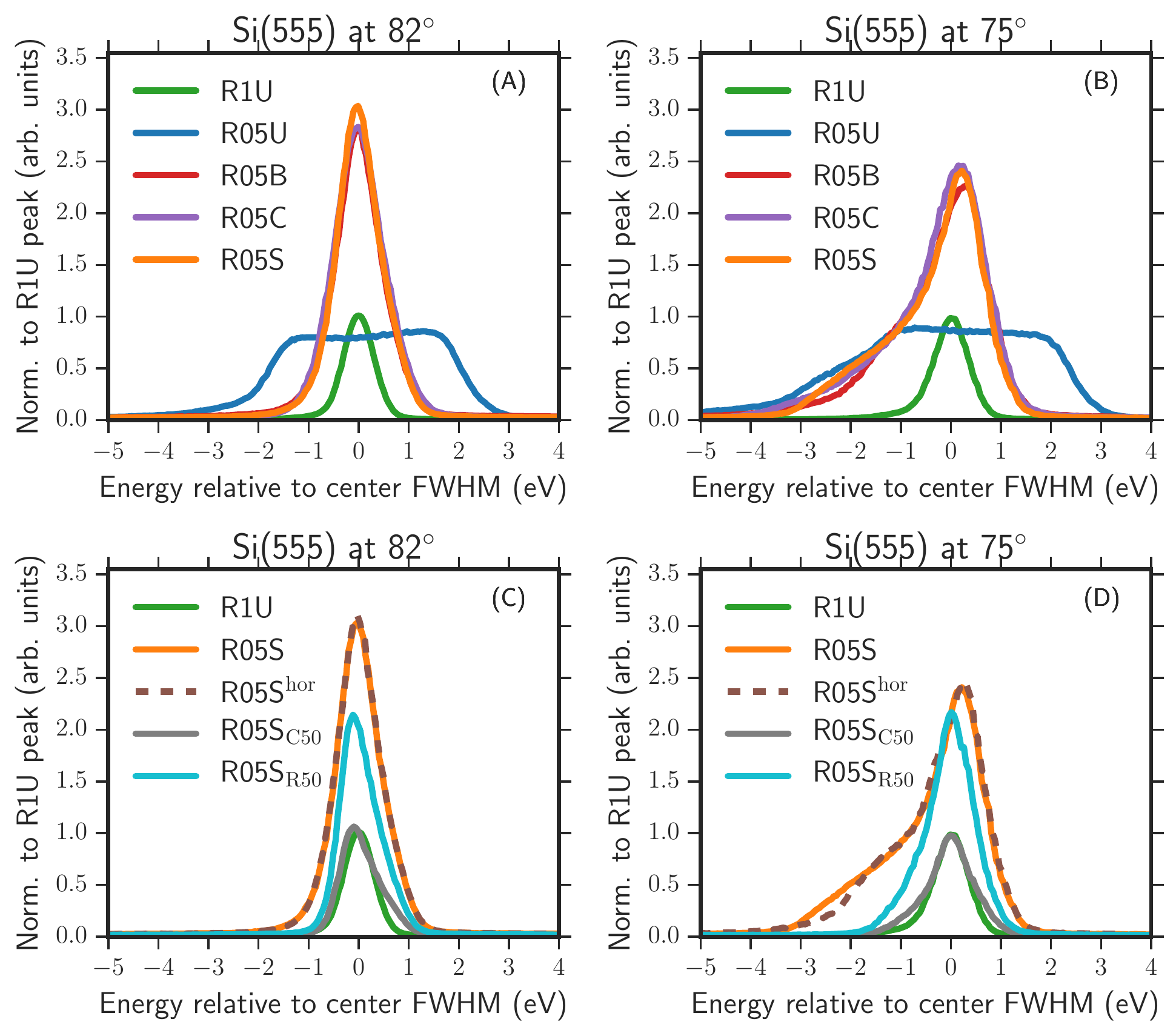}
  \caption{(Color online) Elastic peak scans for the four 0.5~m SBCAs of Fig.~\ref{fig:tested} (R05*) and the 1~m SBCA reference (R1U), collected using the Si(311) incoming beam monochromator. The panels are, respectively, for $\theta_{\rm B}$ 82$^\circ$ (left) and 75$^\circ$ (right). The intensities are normalized to I$_{\rm peak}$ of R1U, as reported in Table~\ref{tab:results}. The abscissa and ordinate plot limits are kept equal between pa\-nels to facilitate visual comparison. Bottom panels (C and D) show R05S performance with respect to 90$^\circ$ rotation of the strips (R05S$^{\rm hor}$) and the use of masks (R05S$_{\rm C50}$ and R05S$_{\rm R50}$).}
  \label{fig:res_main05m}
\end{figure}

\begin{figure}[!htb]
  \centering
  \includegraphics[width=0.49\textwidth]{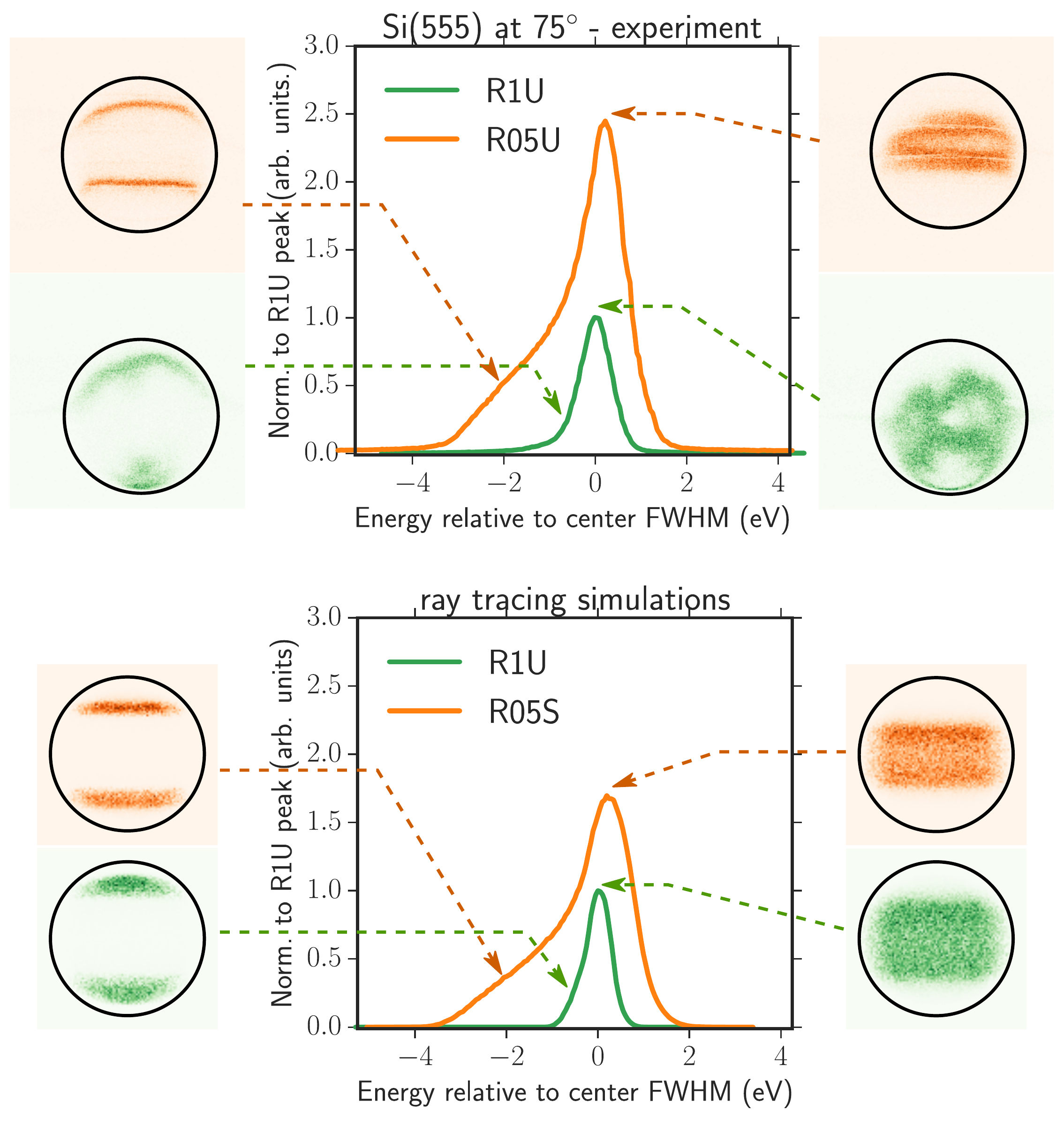}
  \caption{(Color online) Effectively scattering area for R1U and R05S at 75$^\circ$ for two positions on the elastic peak. The black circles represent the contour of the 100~mm crystal analyzer, obtained by summing the images at all energies. For each image, the color gradient is normalized between 0 and the maximum of intensity. The top and bottom panels show experimental data and ray tracing simulations, respectively.}
  \label{fig:res_scattering}
\end{figure}

To validate the strip-bent method, the first question to answer is how the four 0.5~m SBCAs of Fig.~\ref{fig:tested} compare to each other. The results for the Si(555) reflection at 82$^\circ$ and 75$^\circ$ - central energies of 9982~eV and 10234~eV, respectively - are summarized in Table~\ref{tab:results} and shown in Fig.~\ref{fig:res_main05m}. The uniform analyzer (R05U) has an energy bandwidth $>$3.9~eV. This is caused by the high strain in the crystal. The dominant contribution is the angular compression, as demonstrated by the dynamical diffraction simulations results, $\Delta$E$_{\rm TT}^{\rm ang}$ reported in Table~\ref{tab:rt}. When the angular strain is released, either by uncontrolled cracks (R05B and R05C) or by controlled strips (R05S), the experimental energy bandwidth reduces to $\approx$1~eV, as obtained also from the simulation of the vertical component only of the strain field, $\Delta$E$_{\rm ML}$ or $\Delta$E$_{\rm TT}^{\rm v}$ in Table~\ref{tab:rt}. In addition to the peak narrowing, the peak intensity is increased. These results are valid both at 82$^\circ$ and 75$^\circ$. At 75$^\circ$, as expected from the ray tracing simulations - $\Delta$E$_{\rm RT}$ in Table~\ref{tab:rt} and Fig.~\ref{fig:res_scattering} - the low energy tail, mainly attributed to the geometric contributions, degrades both the energy bandwidth and the gain in intensity. In summary, it is found that the angular strain is the dominant contribution in the uniform 0.5~m SBCAs and is strongly reduced by the strip-bent method. The strip-bent 0.5~m SBCAs are limited in energy resolution mainly by the geometric contributions, as found from the ray tracing simulations.

\begin{table}[!htb]
  \caption{Results of the experimental characterization using a Si(311) incoming beam monochromator and for circular Si(111) SBCAs of 100~mm diameter and bending radius R. A mask on the crystal surface may be applied (\emph{cf.} main text). The data were collected with the Si(555) reflection at two Bragg angles ($\theta_{\rm B}$). The strips of R05S were oriented parallel to the Rowland circle (vertically). The energy bandwidth $\Delta$E$_{\rm EXP}$ is given as full width at half maximum of the fitted elastic peaks. The intensities are normalized to the 1~m SBCA reference, R1U, and are reported as peak height and area ratios, I$_{\rm peak}$ and I$_{\rm area}$, respectively. The error bar was estimated from the experimental systematic errors and is reported in parenthesis on the last digit.}
  \label{tab:results}
  \begin{ruledtabular}
    \begin{tabular}{llllcccc}
      ID & Type & R    & Mask & $\theta_{\rm B}$  & $\Delta$E$_{\rm EXP}$ & I$_{\rm peak}$   & I$_{\rm area}$   \\
         &      & (m)  &      & ($^\circ$)  & (eV)                 & \multicolumn{2}{c}{(arb. units)} \\
      \hline
      R1U  & uniform    & 1   & -   & 82 &  0.71(5)  & 1.0(1) & 1.0(1) \\ 
      R05U & uniform    & 0.5 & -   & 82 &  3.93(5)  & 0.8(1) & 4.1(1) \\ 
      R05B & cracked    & 0.5 & -   & 82 &  1.05(5)  & 2.7(1) & 4.0(1) \\ 
      R05C & cracked    & 0.5 & -   & 82 &  1.15(5)  & 2.8(1) & 4.3(1) \\ 
      R05S & strip-bent & 0.5 & -   & 82 &  1.01(5)  & 3.0(1) & 4.1(1) \\
      R05S & strip-bent & 0.5 & C50 & 82 &  0.88(5)  & 1.0(1) & 1.2(1) \\
      R05S & strip-bent & 0.5 & R50 & 82 &  0.88(5)  & 2.0(1) & 2.5(1) \\
      \hline
      R1U  & uniform    & 1   & -   & 75 &  0.81(5)  & 1.0(1) & 1.0(1) \\  
      R05U & uniform    & 0.5 & -   & 75 &  4.77(5)  & 0.9(1) & 4.6(1) \\ 
      R05B & cracked    & 0.5 & -   & 75 &  1.65(5)  & 2.3(1) & 4.5(1) \\ 
      R05C & cracked    & 0.5 & -   & 75 &  1.55(5)  & 2.5(1) & 4.9(1) \\ 
      R05S & strip-bent & 0.5 & -   & 75 &  1.51(5)  & 2.4(1) & 4.5(1) \\
      R05S & strip-bent & 0.5 & C50 & 75 &  0.99(5)  & 1.0(1) & 1.2(1) \\
      R05S & strip-bent & 0.5 & R50 & 75 &  1.07(5)  & 2.1(1) & 2.7(1) \\
    \end{tabular}
  \end{ruledtabular}
\end{table}

\begin{table}[!htb]
  \caption{Results of the dynamical diffraction and ray tracing si\-mu\-la\-tions for two circular Si(111) SBCAs of 100~mm diameter and bending radius R. The Bragg angle $\theta_{\rm B}$ and the corresponding nominal energy E for the Si(555) reflection are reported. The contributions to the energy broadening are: $\Delta$E$_{\rm ML}$, $\Delta$E$_{\rm TT}^{\rm v}$, $\Delta$E$_{\rm TT}^{\rm ang}$ and $\Delta$E$_{\rm RT}$, respectively, from the multi-lamellar, Takagi-Taupin vertical and angular plus the ray tracing simulations. $\Delta$E$_{\rm RT}$ includes the diffraction profile calculated with the multi-lamellar method and the geometrical contributions. $\Delta E$, are given as full width at half maximum of the fitted peaks.}
  \label{tab:rt}
  \begin{ruledtabular}
    \begin{tabular}{lllllll}
      R      & $\theta_{\rm B}$     & E     & $\Delta$E$_{\rm ML}$ & $\Delta$E$_{\rm TT}^{\rm v}$ & $\Delta$E$_{\rm TT}^{\rm ang}$ & $\Delta$E$_{\rm RT}$ \\
      (m)   & ($^\circ$)  & (eV)  & (eV)   & (eV)    & (eV)        & (eV) \\
      \hline 
       1   & 82 &  9982  &  0.20 & 0.24 & 0.98 & 0.46  \\ 
       0.5 & 82 &  9982  &  0.41 & 0.48 & 3.93 & 0.89  \\ 
       \hline                         
       1   & 75 &  10234 &  0.18 & 0.26 & 1.00 & 0.74  \\  
       0.5 & 75 &  10234 &  0.35 & 0.51 & 4.03 & 1.49  \\ 
    \end{tabular}
  \end{ruledtabular}
\end{table}

The orientation of the strips with respect to the Rowland circle (scattering plane) was also investigated in order to determine if it affects the energy resolution of the SBCAs. In fact, the strips could introduce an asymmetry in the strain field of the crystal or a figure error if the bonding is not perfect. We found that neither the energy bandwidth nor the scattered intensity are modified, within a 5\% variance, if the strips are oriented parallel or perpendicular to the Rowland circle. The test was performed on the Si(111) (Fig.~\ref{fig:res_main05m}C and \ref{fig:res_main05m}D) and the Si(110), Si(311), Si(951), Ge(111), Ge(110) and Ge(100) (not shown) orientations. 

The following step consists in comparing 0.5~m with 1~m SBCAs in terms of energy bandwidth and scattered intensity. The results are also shown in Fig.~\ref{fig:res_main05m} and reported in Table~\ref{tab:results}. For the 1~m SBCAs, we did not observe significant differences between a uniform and a strip-bent crystal analyzer (results not shown). We chose a uniform 1~m SBCA (R1U) because this type is currently the most used. We underline that R1U is of high quality and its performances identical to those for a strip-bent or a bent-diced analyzer with stress-relief cuts. A standard quality 1~m SBCA has an energy bandwidth of $\approx$1~eV in the same experimental conditions. For the 0.5~m strip-bent SBCA, R05S, we found that the energy bandwidth is comparable to that of the 1~m SBCA, R1U, at 82$^\circ$, whereas the peak maximum increased by a factor $\approx$3 and the peak area by a factor $\approx$4. These values decrease at 75$^\circ$ because the geo\-me\-tri\-cal contributions degrade the performance of the 0.5~m SBCAs as a function of $\theta_{\rm B}$ faster than that of 1~m SBCAs. A gain factor in the peak area ratio higher than four, given by the increase in the solid angle, is explained by the broader reflectivity of the crystal introduced by the higher strain at 0.5~m. The experimental results are confirmed by the ray tracing simulations where the area ratio is 4.4(1) for both angles.

The low energy tail that degrades the energy bandwidth at 75$^\circ$ can be reduced by masking the surface of the analyzer. This was tested by using a circular mask of 50~mm diameter (C50) and a rectangular one of 100$\times$50~mm$^2$ (R50), in which the shorter side was oriented along the Rowland circle (vertical). The results for RS05S are reported in Table~\ref{tab:results} and shown in Fig.~\ref{fig:res_main05m}. The circular mask allows reducing the energy bandwidth down to the one of R1U at the expense of intensity. In contrast, the rectangular mask results in a reduced energy bandwidth with an intensity gain of a factor $\approx$2/2.5 in the peak/area ratios. Using a rectangular mask may proof advantageous in applications where the energy resolution is crucial.

\begin{figure}[!htb]
  \centering
  \includegraphics[width=0.49\textwidth]{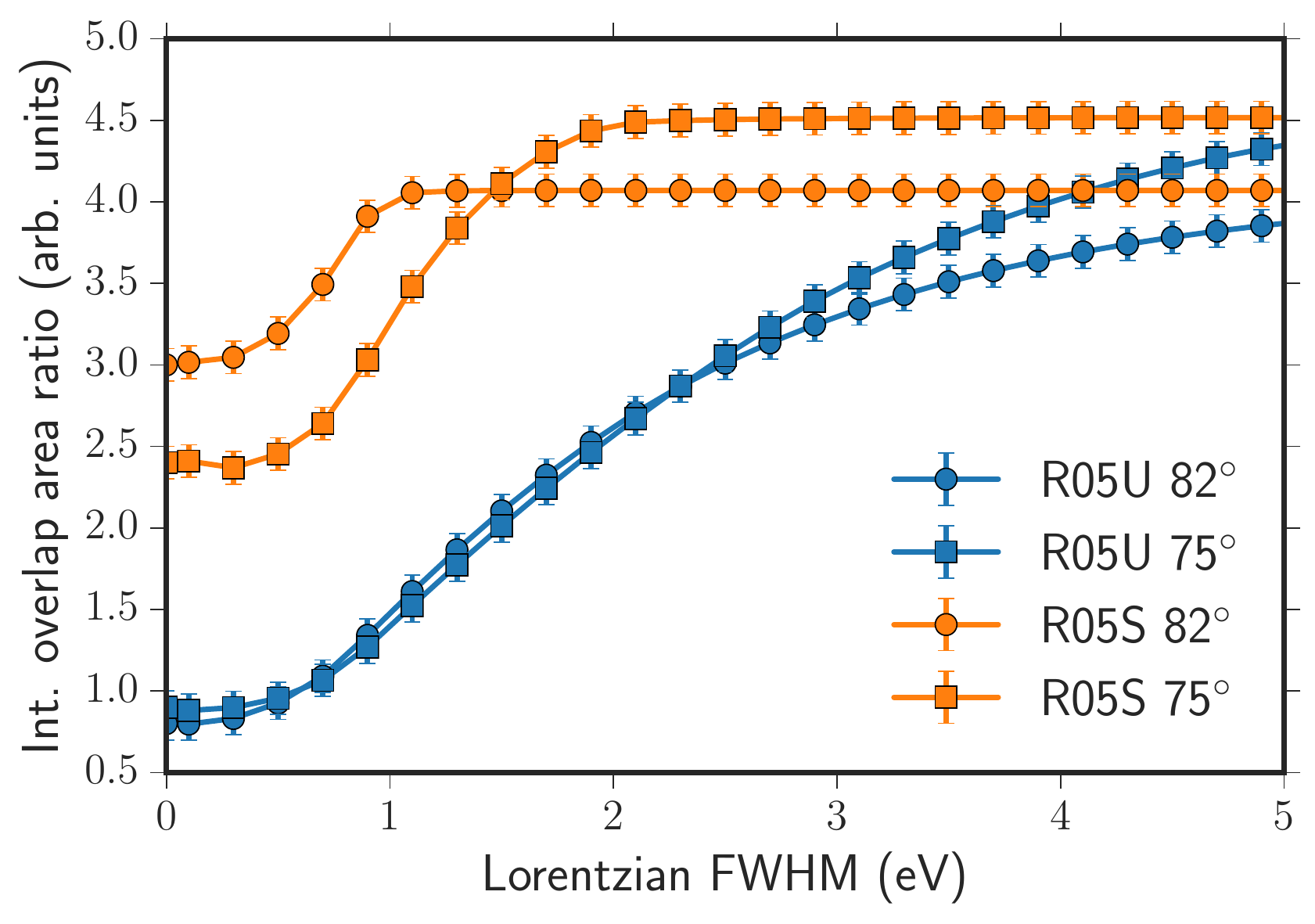}
  \caption{(Color online) Evolution of the intensity ratio between 0.5~m and 1~m SBCAs as a function of the FWHM of a Lorentzian function (\emph{cf.} main text). For clarity, only R05U/R1U and R05S/R1U are shown.}
  \label{fig:res_intratio}
\end{figure}

The intensity gain between 0.5~m and 1~m SBCAs depends on the natural width of the measured emission line for a given $\theta_{\rm B}$. In Fig.~\ref{fig:res_intratio} it is shown how the intensity gain for the strip-bent (R05S) and the uniform (R05U) evolves as a function of the natural broadening of an emission line. The graph was obtained by integrating the area that overlaps the experimental elastic peaks of Fig.~\ref{fig:res_main05m} and a Lorentzian function (representing the emission line) normalized to the peak height and centered on the experimental elastic peaks. The ordinate in Fig.~\ref{fig:res_intratio} shows this overlap area for the 0.5~m crystal analyzers normalized to the value obtained for the 1~m crystal analyzer. The FWHM of the Lorentzian function was varied in the range 0.1--5~eV. For a sharp emission line, approaching a $\delta$ function, the intensity gain is given by the peak ratio, whereas when the emission line is broad, $\ge$5~eV, the gain is given by the area ratio. This is an important point to take into account when planning an experiment.

The origin of such difference is obtained by investigating how the area of the analyzers contributes to the scattered intensity. The scattering surface can be imaged by moving a position sensitive detector out of the Rowland circle. If the detector is in between the meridional and sagittal focus, respectively, $\rm q_{m}$ and $\rm q_{s}$ image distances, the focal spot is circular. Using Coddington's equations the detector offset in the direction of the reflected beam is $\rm (q_{s}-q_{m})/2$, where $\rm q_{m}=R \cdot sin(\theta_B)$, $\rm q_{s}=R\cdot\sin(\theta_B)/(2\cdot\sin^2(\theta_B)-1)$ and R is the bending radius of the SBCA. In our experiment the detector offset at 75$^\circ$ was $\approx$37~mm and $\approx$75~mm for the 0.5~m and 1~m configurations, respectively. Fig.~\ref{fig:res_scattering} shows how the R05S and R1U effectively scatter at 75$^\circ$, for two points on the elastic peak: 1) at the peak low energy shoulder and 2) at the peak maximum. In the first position only the external parts of the analyzer in the vertical direction contribute to the scattered intensity, due to the geometric aberrations. At the peak maximum mainly the central part of the analyzer reflects. These results are confirmed by the ray tracing simulations shown in the bottom panels of Fig.~\ref{fig:res_scattering}. In summary, to further improve the performances of the 0.5~m SBCAs it is crucial to correct the geometrical aberrations. This could be achieved by manufacturing a Johansson-type \cite{Johansson:1933_ZP} toroidal analyzer, as proposed by Wittry and Sun with a stepped surface \cite{Wittry:1991_JAP}.

\subsection{Example of application}
\label{sec:exp_application}

\begin{figure}[!htb]
  \centering
  \includegraphics[width=0.49\textwidth]{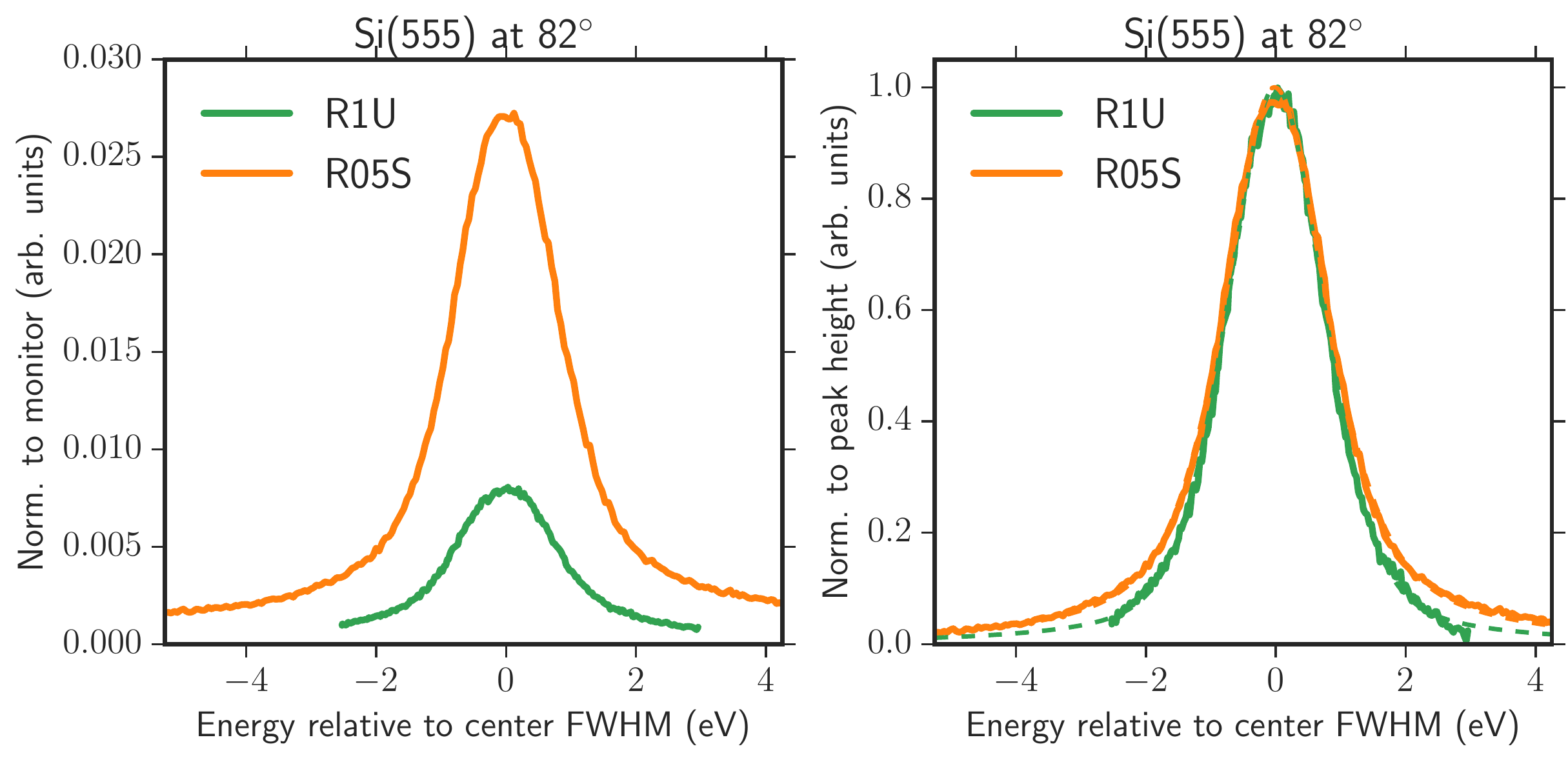}
  \caption{(Color online) Intensity and energy resolution comparison between 0.5~m strip-bent SBCA (R05S) and 1~m uniform SBCA (R1U) with a Si(111) incoming beam monochromator at 9982~eV ($\theta_{\rm B}$ of 82$^\circ$). The left panel shows the data normalized to the incoming beam flux. The right panel shows the data normalized to the peak height of the fitted asymmetric Pseudo-Voigt profiles, where a constant background has been subtracted. The dashed lines show the fitted profiles.}
  \label{fig:res_mono111}
\end{figure}

For applications requiring high photon flux, the energy bandwidth of the incoming x-rays is usually increased to gain in intensity. For example, by employing a Si(111) double crystal monochromator a factor five in flux may be gained with respect to a double Si(311) at the same energy. We found that with a Si(111) incoming beam monochromator, the energy bandwidth of the Si(555) reflection at 82$^\circ$ is equivalent between 0.5~m and 1~m SBCAs, being 1.93(5)~eV for R05S and 1.80(5)~eV for R1U. In contrast, the gain in intensity of R05S with respect to R1U is 3.6(1) and 4.2(1) in the peak and area ratio, respectively. These results are shown in Fig.~\ref{fig:res_mono111}.

\begin{figure}[!htb]
  \includegraphics[width=0.49\textwidth]{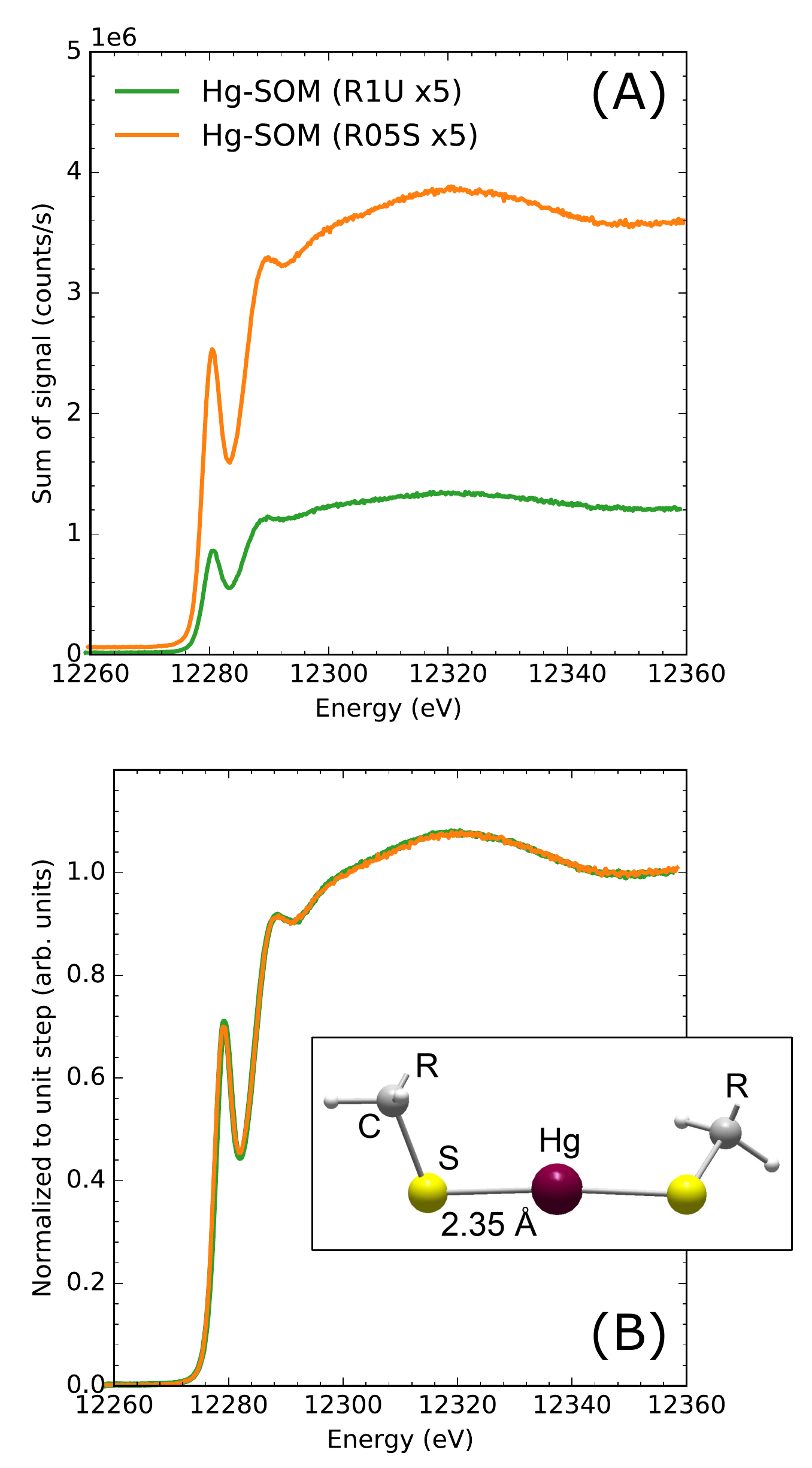}
  \caption{(Color online) Hg L$_3$-edge HERFD-XANES spectra of Hg-SOM collected at the maximum of the Hg L$\alpha_1$ line with five SBCAs at 0.5~m (orange) and 1~m (green) bending radius. The top panel (A) shows the intensity gain obtained for the same counting time. The bottom panel (B) shows that the normalized spectra have identical resolution. The inset shows a ball-and-stick model of the coordination of Hg in SOM.}
  \label{fig:res_xanes}
\end{figure}

An example of application in environmental science is presented in Fig.~\ref{fig:res_xanes}. The sample is a soil organic matter (SOM) reacted for 15 hours with 200~ppm (ng/mg) divalent mercury~\cite{Manceau:2015_EnvSciTech} (Hg-SOM). The Hg L$_3$-edge HERFD-XANES (x-ray absorption near-edge structure) spectra were collected at the maximum of the Hg L$\alpha_1$ line (L$_3$-M$_5$) in two experimental sessions with the spectrometer equipped with a set of five SBCAs of either 1~m (R1U x5) or 0.5~m (R05S x5) radius. The data were collected at a temperature of $\approx$15~K on the same pellet with scans of 15~s duration. Each scan was collected on a fresh spot to minimize radiation damage. Ninety scans were measured in total at each session and summed to improve the counting statistics (Fig.~\ref{fig:res_xanes}, panel A). The gain effectively obtained with the 0.5~m SBCAs was 2.9(1), for an expected value of 4.2(1) based on the area ratio of the elastic peaks of Fig.~\ref{fig:res_mono111} and considering that the width of the Hg L$\alpha_1$ line is 2.28 eV \cite{Schoonjans:2011_SAB}, larger than the crystal analyzers bandwidth. The difference in gain is explained by the fact that during the measurements the signal from two outermost R05S crystal analyzers was partially blocked by the mechanical mount in front of the detector and the sample inhomogeneity.

The sample was chosen because its spectrum has a sharp near-edge peak at 12279~eV, which can be used to compare the spectral resolutions provided by the 1~m and 0.5~m SBCAs. This peak results from the electronic transition from 2$p_{3/2}$ to 5$d$ and 6$s$ mixed states and is diagnostic of Hg coordinated linearly to two thiol groups \cite{Manceau:2015_EnvSciTech,Manceau:2015_IC}. As shown in the bottom panel (B) of Fig.~\ref{fig:res_xanes}, the two normalized spectra are identical. This result opens the way for establishing high energy-resolution spectroscopy at ultra-high dilution. Recently \cite{Manceau:2016_EST}, good quality Hg L$_3$-edge HERFD-XANES spectra were obtained at concentrations as low as 2~ppm and 0.5~ppm Hg in human hair after two and eight hours of acquisition time, respectively. A highlight of this study has been the detection of amine ligands at 2.5-2.6~\AA\ in the coordination sphere of Hg which give a distinctive spectral feature only seen at high energy-resolution.

\section{Conclusions and Outlook}
\label{sec:outlook}

We have presented the manufacturing and performance of strip-bent SBCAs at 0.5~m ben\-ding radius. These crystal analyzers are a powerful alternative to the commonly used 1~m SBCAs when dealing with (ultra-) dilute or radiation sensitive species. Strip-bent 0.5~m SBCAs collect more photons and have an energy resolution close to the limit given by intrinsic and geometrical contributions. The gain in intensity with respect to 1~m SBCAs ranges from 2.5 to 4.5, depending on the width of the detected emission line. We have demonstrated that for applications requiring high flux the possible gain at $\approx$10~keV is between 3.6 and 4.2 without performance trade-off in energy resolution. To date, the high-luminosity analyzers described herein have been successfully used to speciate Hg at natural dilution down to 0.5~ppm. This example establishes the foundation for applying the new instrumentation to other systems of metals (\emph{e.g.} Au, Ag, Pt), metalloids (\emph{e.g.}, As, Se), rare-earth elements, and radionuclides, in biological, chemical, environmental, Earth, materials and forensic science. Beyond their immediate utility at synchrotron radiation facilities, 0.5~m SBCAs open an avenue for the development of high energy-resolution spectroscopies in the laboratory.

\begin{acknowledgments}
M.R. would like to acknowledge Manuel Sanchez del Rio for fruitful discussions on the ray tracing simulations. This work was supported by the French National Research Agency (ANR) under Grant ANR-10-EQPX-27-01 (EcoX Equipex).\\
\end{acknowledgments}

\bibliographystyle{apsrev4-1}

\end{document}